\documentclass[prb,floatfix,showpacs,amssymb]{revtex4}

\usepackage{graphicx}
\usepackage{dcolumn}
\usepackage{bm}

\begin{document}
\title{Long range statistical fluctuations of the crossed Josephson current}
\author{R\'egis M\'elin\footnote{regis.melin@grenoble.cnrs.fr}}
\affiliation{
Centre de Recherches sur les Tr\`es Basses
Temp\'eratures (CRTBT\footnote{U.P.R. 5001 du CNRS, Laboratoire
conventionn\'e avec l'Universit\'e Joseph Fourier}),\\ CNRS, BP 166,
38042 Grenoble Cedex 9, France}

\begin{abstract}
We investigate the crossed
Josephson effect in a geometry consisting of a 
double ferromagnetic bridge between two superconductors,
with tunnel interfaces. The
crossed Josephson current vanishes on average because the
Andreev reflected hole does not follow the same sequence of
impurities as the incoming electron.
We show that i) the root mean square of the 
crossed Josephson current distribution is proportional to the
square root of the junction area; and ii) the coherent coupling
mediated by fluctuations is ``long range'' since it decays
over the ferromagnet phase coherence length $l_\varphi$, larger than the
exchange length. We predict a
crossed Josephson current due to fluctuations if the length 
of the ferromagnets is smaller than $l_\varphi$ and larger than
the exchange length~$\xi_h$. 
\end{abstract}

\maketitle

\section{Introduction}

Transport properties of hybrid structures consisting of
a superconductor (S) multiply connected to several normal metal (N)
or ferromagnetic (F) electrodes has focused an important interest
recently\cite{Allsop94,Lambert98,Jedema99}. In usual Andreev reflection
at a single NS interface,
a spin-up electron coming from the N side is reflected as a 
hole in the spin-down band while a Cooper pair is transferred in the
superconductor. Multiterminal structures allow ``non local'' processes,
in which a spin-up electron in one electrode is Andreev reflected
as a hole in the spin-down band
in another electrode, corresponding to non
local transmission in the electron-hole channel. Conversely
non local transmission in the electron-electron channel
corresponds to a process
in which a spin-up electron from one electrode is transmitted
as a spin-up electron in another electrode. Transport theory 
of three-terminal FSF junctions
including non local transmission in the electron-electron and
electron-hole channels has been discussed
recently\cite{Deutscher00,Falci01,Melin01,Melin02,Tadei02,Melin03,Chte03,AB03,Feinberg03,Sanchez03,Lambert03,Yamashita03,Bignon04,Melin04,Prada04},
in the tunnel limit\cite{Falci01,Prada04} and for highly transparent
interfaces\cite{Melin02,Melin04},
on the basis of microscopic Green's
functions\cite{Falci01,Melin02,Melin04,Melin01,Melin03,AB03}
and in the framework
of the scattering approach\cite{Deutscher00,Sanchez03,Yamashita03}.
The models were also
extended to
describe disorder\cite{Chte03,Feinberg03,Lambert03},
non collinear ferromagnets\cite{Melin03}, and the
noise\cite{Tadei02,Bignon04}. On the experimental side, two experiments
probing non local transport
were carried out recently\cite{Beckmann04,Russo05}, in FSF and NSN
three-terminal junctions.

\begin{figure}
\includegraphics [width=.6 \linewidth]{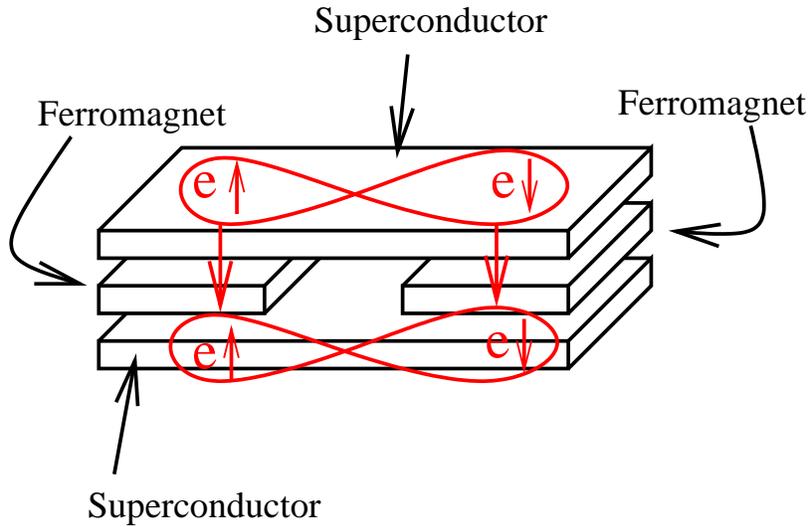}
\caption{(Color online.) Schematic 3D representation of the Josephson junction
considered in the article. 
\label{fig:schema-3D}
}
\end{figure}

The question arises of whether the phase coherence of crossed
Andreev reflection can be probed experimentally. We show here that this
is possible with a
double ferromagnetic bridge between two superconductors
(see Fig.~\ref{fig:schema-3D}).
We have already shown that the crossed Josephson current vanishes on
average in the diffusive limit\cite{Melin03}, because the Andreev reflected
hole does not follow the same sequence of impurities as the incoming
electron since they propagate in different electrodes. However, by evaluating
the statistical
fluctuations of the dc crossed Josephson current, we show here that
the fluctuations of the Josephson current
decay over the phase coherence length $l_\varphi$ in the ferromagnet,
larger than the decay length of the
local average Josephson current set by the exchange length
$\xi_h$ (see
Refs.~\onlinecite{Buzdin,Radovic,Heikkila,Golubov-short,osc,Aarts,Rya,Kontos,Guichard,Sellier}).
The fluctuations of the crossed supercurrent do not show
$\pi$-shift oscillations and damping as a function of the ferromagnet
length, because the spin-up and spin-down
electrons of correlated pairs extracted from one superconductor do not see the
same realization of disorder, so that the center of mass momentum of
the spatially separated
correlated pair averages to zero after propagation over a length comparable to
the elastic mean free path. 
The root mean square of the crossed
Josephson current
is proportional to the
square root of the junction area, because the number of diagrams
involved in the supercurrent is equal to the junction area divided 
by the Fermi wave-length.
The crossed Josephson supercurrent due to fluctuations
can in principle
be detected experimentally, provided
the length of the ferromagnets is smaller than $l_\varphi$
and larger than $\xi_h$.

The article is organized as follows.
Preliminaries regarding
Green's functions are given in
section~\ref{sec:prelim}. The analytical results are presented in
section~\ref{sec:av-loc-super} for the average local supercurrent,
and in section \ref{sec:fluc-super} for the statistical
fluctuations of the
supercurrent. 
Concluding remarks are given in
section~\ref{sec:conclu}.
Some details on disorder averaging are provided in the Appendix.

\begin{figure}
\includegraphics [width=.6 \linewidth]{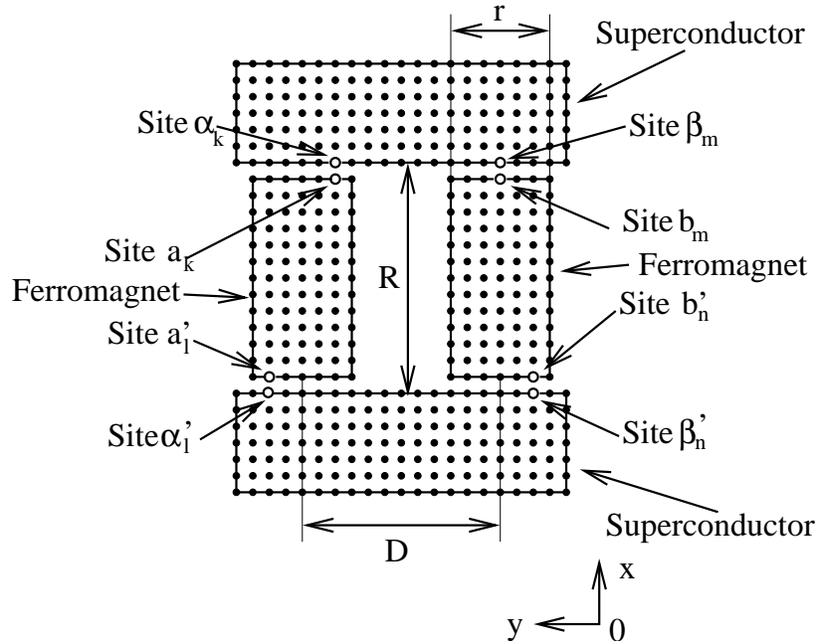}
\caption{Schematic 2D cut of the junction on Fig.~\ref{fig:schema-3D}.
We have represented some pairs of sites at the interfaces:
$(a_k,\alpha_k)$, $(a'_l,\alpha'_l)$, $(b_m,\beta_m)$, and
$(b'_n,\beta'_n)$. 
\label{fig:schema-coupe}
}
\end{figure}

\section{Technical preliminaries}
\label{sec:prelim}
\subsection{The models}
The superconductor is described by the BCS Hamiltonian
\begin{equation}
{\cal H}_{\rm BCS} = \sum_{\langle \alpha , \beta
\rangle , \sigma} - t \left(
c_{\alpha,\sigma}^+ c_{\beta,\sigma}
+c_{\beta,\sigma}^+ c_{\alpha,\sigma} \right)
+ \Delta \sum_\alpha \left(
c_{\alpha,\uparrow}^+ c_{\alpha,\downarrow}^+
+ c_{\alpha,\downarrow}
c_{\alpha,\uparrow} \right)
,
\end{equation}
where $t$ is the hopping amplitude,
$\Delta$ is the superconducting gap, and
$\alpha$ and $\beta$ correspond to neighboring sites on a
cubic lattice with a parameter $a_0$. The lattice parameter
$a_0$ is
taken equal to the Fermi wave-length $\lambda_F$.
The ferromagnetic electrodes are described by the Stoner model
\begin{equation}
\label{eq:Stoner}
{\cal H}_{\rm Stoner} = \sum_{\langle \alpha,\beta
\rangle,\sigma} -t \left( c_{\alpha,\sigma}^+
c_{\beta,\sigma} + c_{\beta,\sigma}^+ 
c_{\alpha,\sigma} \right)
-h_{\rm ex} \sum_\alpha
\left( c_{\alpha,\uparrow}^+ c_{\alpha,\uparrow}
-c_{\alpha,\downarrow}^+ c_{\alpha,\downarrow} \right)
,
\end{equation}
where $h_{\rm ex}$ is the exchange field. The exchange fields
in the two ferromagnets are equal in the
parallel alignment, and opposite in the antiparallel alignment.
We incorporate also disorder scattering,
described by the Hamiltonian
\begin{equation}
\label{eq:H-dis}
{\cal H}_{\rm dis}=\sum_{\alpha_n,\sigma} V_{\alpha_n}
c_{\alpha_n,\sigma}^+ c_{\alpha_n,\sigma}
,
\end{equation}
where the impurities are located at the sites $\alpha_n$. The
impurity scattering potentials $V_{\alpha_n}$ are random variables.
The site $a_k$ is on the ferromagnetic
side of the interface, and the site $\alpha_k$ on the superconducting side.

The couplings between the ferromagnets and the superconductors are
described
by the tunnel Hamiltonian. 
The tunnel Hamiltonian at the interface
$(a,\alpha)$ takes the form
\begin{equation}
\label{eq:tunnel-ab}
{\cal W}_{a,\alpha} = \sum_{k,\sigma} \left( - t_{a_k,\alpha_k} 
c_{a_k,\sigma,F}^+ c_{\alpha_k,\sigma,S} - t_{\alpha_k,a_k} 
c_{\alpha_k,\sigma,S}^+ c_{a_k,\sigma,F} \right) 
,
\end{equation}
where the summation runs over all sites at the interface (see
Fig.~\ref{fig:schema-coupe}), and where $t_{a_k,\alpha_k}
=t_{\alpha_k,a_k}$ is the hopping amplitude connecting the sites
$a_k$ and $\alpha_k$.

\subsection{Green's functions of a ferromagnet and a superconductor}
The starting point is the ballistic
Green's function $\hat{g}_{i,j}(\omega)$
of the isolated ferromagnetic and
superconducting electrodes in the Nambu representation.
The ballistic Green's functions of a  ferromagnet take the form
\begin{eqnarray}
\label{eq:g11-ferro}
g_{a,b}^{\uparrow,1,A}(\omega)&=&-\frac{\pi \rho_F}{k_F d_{a,b}}
\exp{\left[-i \left(k_F^\uparrow + \frac{\omega}{\hbar v_F^\uparrow}
\right) d_{a,b} \right]} \exp{(-d_{a,b}/l_\varphi^{({\rm ball})})}\\
g_{a,b}^{\uparrow,2,A}(\omega)&=&\frac{\pi \rho_F}{k_F d_{a,b}}
\exp{\left[i \left(k_F^\downarrow - \frac{\omega}{\hbar v_F^\downarrow}
\right) d_{a,b} \right]} \exp{(-d_{a,b}/l_\varphi^{({\rm ball})})}
\label{eq:g22-ferro}
,
\end{eqnarray}
where $g_{a,b}^{(\uparrow,1)}$ and $g_{a,b}^{(\uparrow,2)}$ are 
the Green's functions of a spin-up electron and a hole
in the spin-down band respectively, both having $S_z=1/2$,
$d_{a,b}$ is the distance between the sites $a$ and $b$,
$\omega$ the energy with respect to the chemical potential,
$\rho_F$ the density of states, $k_F^\uparrow$ and
$k_F^\downarrow$ the spin-up and spin-down Fermi wave-vectors,
$v_F^\uparrow$ and $v_F^\downarrow$ the spin-up and spin-down Fermi
velocities. 
The ballistic ferromagnet Green's functions given by Eqs.~(\ref{eq:g11-ferro})
and (\ref{eq:g22-ferro}) decay exponentially over the phase coherence 
length $l_\varphi^{({\rm ball})}$, introduced 
phenomenologically through an
imaginary part $\hbar v_F/l_\varphi^{({\rm ball})}$ to the energy $\omega$.
We note $k_F$ and $v_F$ the Fermi wave-vector and the Fermi velocity
in the absence of spin polarization.
We neglect in the following the energy dependence of the ferromagnet
propagators in Eqs.~(\ref{eq:g11-ferro}) and (\ref{eq:g22-ferro})
since we suppose that the length $R$ of the ferromagnets
is small compared $\hbar v_F^\uparrow/\Delta$ and $\hbar
v_F^\downarrow/\Delta$, both length scales being comparable to the
ballistic BCS coherence length $\hbar v_F/\Delta$.

The Nambu Green's function of a
ballistic isolated superconductor in the sector $S_z=1/2$ takes the form
\begin{equation}
\hat{g}_{\alpha,\beta}(\omega)=
\frac{\pi \rho_S}{k_F d_{\alpha,\beta}} \exp{
\left(-\frac{d_{\alpha,\beta}}{\xi_{\rm BCS}^{({\rm ball})}
(\omega)}\right)}
\left\{ \frac{\sin{(k_F d_{\alpha,\beta})}}{\sqrt{\Delta^2-\omega^2}}
\left[ \begin{array}{cc} -\omega & \Delta \\ \Delta & -\omega
\end{array} \right] 
+\cos{(k_F d_{\alpha,\beta})}
\left[ \begin{array}{cc} -1 & 0 \\ 0 & 1 \end{array} \right]
\right\}
,
\label{eq:g-nonloc-S}
\end{equation}
where $\rho_S$ is the normal state density of states of the
superconductor,
$d_{\alpha,\beta}$
the distance between the sites $\alpha$ and $\beta$,
and $\xi_{\rm BCS}^{({\rm ball})}(\omega)=
\hbar v_F/\sqrt{\Delta^2-\omega^2}$ the
BCS coherence length at a finite energy. The information about
propagation in the superconductor in the non local Josephson effect
is contained in $f_{\alpha,\beta}(\omega)\equiv
g^{1,2}_{\alpha,\beta}(\omega)$, where ``1'' and ``2'' refer to the
electron and hole Nambu labels respectively.
The statistical fluctuations
of the Josephson current involve $\overline{\left(f_{\alpha,\beta}(\omega)
\right)^2}$, where the overline is an average over disorder and over
the different conduction channels. We have\cite{Feinberg-des,Smith}
\begin{equation}
\label{eq:avf2}
\overline{\left(f_{\alpha,\beta}(\omega)\right)^2}
=\frac{\pi \rho_S}{k_F^2 l_d d_{\alpha,\beta}} \frac{\Delta^2}
{\Delta^2-\omega^2}
\exp{\left(-\frac{d_{\alpha,\beta}}{\xi_{\rm BCS}(\omega)}\right)}
,
\end{equation}
where $\xi_{\rm BCS}(\omega)$
is the diffusive limit superconducting coherence length.
\subsection{Supercurrent}
\begin{figure*}
\includegraphics [width=1. \linewidth]{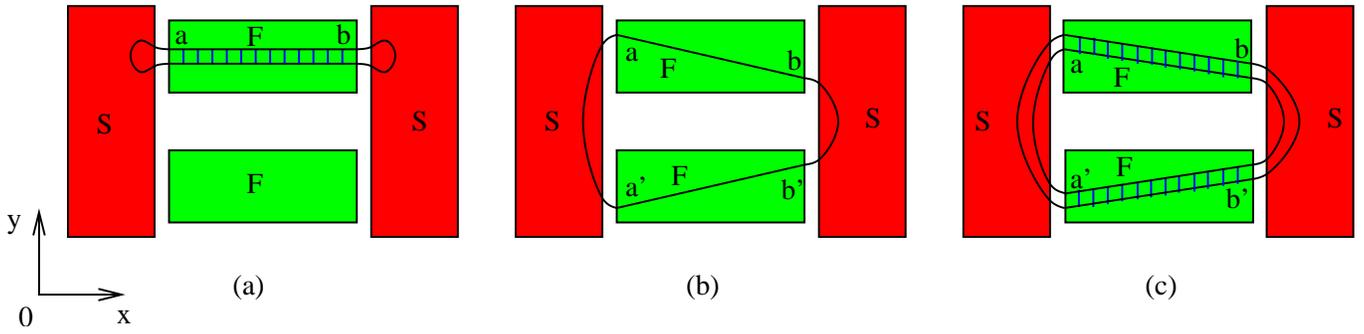}
\caption{(Color online.) 
Schematic representation of the lowest order diagrams for
the ``local'' supercurrent (a), the ``non local'' supercurrent (b),
and the statistical fluctuations of the ``non local'' supercurrent (c).  
We represent schematically the ladder diagrams for the diffuson.
The rungs of the ladders correspond to a disorder scattering.
\label{fig:diag}
}
\end{figure*}

The fully dressed Green's functions $\hat{G}_{i,j}(\omega)$ are obtained
from the Dyson equation $\hat{G}(\omega)=\hat{g}(\omega)+
\hat{g}(\omega)\otimes\hat{\Sigma}
\otimes \hat{G}(\omega)$, where $\otimes$ corresponds to
a summation over all the sites
in the tunnel Hamiltonian (\ref{eq:tunnel-ab}). 
The self-energy is provided by the couplings of the tunnel Hamiltonian,
that, in the Nambu representation, take the form
\begin{equation}
\hat{t}_{a,\alpha} = \left[ \begin{array}{cc}
t_a \exp{(i\varphi/4)} & 0 \\
0 & -t_a \exp{(-i\varphi/4)} \end{array} \right]
,
\end{equation} 
where $\varphi$ is the phase difference between the two superconductors
and $t_a$ is a real number.
The phase does not evolve in time since we restrict here to
the dc-Josephson effect.
The second order diagrams on Fig.~\ref{fig:diag} contributing the supercurrent
acquire a phase $\exp{(\pm i \varphi)}$, giving rise to a supercurrent
proportional to $\sin{\varphi}$.
The equilibrium supercurrent through electrode ``a'' is given by
\begin{eqnarray}
\label{eq:sp-I-def}
I_S&=&\frac{e}{h} \int_0^{+\infty}
\mbox{Tr}
\left\{ \hat{\sigma}^z \left[ 
\hat{t}_{\alpha,a} \left( \hat{G}^A_{a,\alpha}(\omega) 
-\hat{G}^R_{a,\alpha}(\omega) \right) 
- \hat{t}_{a,\alpha} \left( \hat{G}^A_{\alpha,a}(\omega) 
-\hat{G}^R_{\alpha,a}(\omega) \right) \right] \right\}
d \omega\\
&+&\left( h_{\rm ex} \rightarrow - h_{\rm ex} \right)
\nonumber
,
\end{eqnarray}
where the trace is a summation over the Nambu labels and the different
conduction channels. 
The term $(h_{\rm ex}\rightarrow-h_{\rm ex})$ corresponds to the
contribution in the sector $S_z=-1/2$.
Eq.~(\ref{eq:sp-I-def}) can be demonstrated from
the expression of the current in terms of the Keldysh Green's
function~\cite{Nozieres,Cuevas}.
The transparency of a single junction in the normal state is proportional
to $(t/\epsilon_F)^2$, where $\epsilon_F$ is the Fermi energy.
We suppose here that $t \ll \epsilon_F$, so
that the supercurrent is expanded
to order $t^4 \rho_S^2 \rho_F^2$. The tunnel
supercurrent coupling coherently
the two ferromagnets is provided by
the emission of a correlated pair of electrons from the left
superconductor by Andreev reflection, followed by the
absorption of the correlated pair by an Andreev reflection
at the right superconductor.
These two Andreev reflections can be ``local''
or ``non local'' in the sense that the incoming electron and
outgoing hole can propagate in identical or in different electrodes.

The local term $I_S^{({\rm loc})}$ in the supercurrent
involves a diagram with
propagation in a single ferromagnet (see Fig.~\ref{fig:diag}-(a)),
such that the incoming electron and the Andreev reflected hole
are scattered by the same sequence of impurities:
\begin{equation}
\label{eq:IS-loc}
I_S^{({\rm loc})}=4\pi \frac{e}{h}\Delta |t_{a,\alpha}|^2
|t_{b,\beta}|^2 (\pi\rho_S)^2 \sin{\varphi}
\mbox{Re} \left\{
\sum_{a,b} 
\overline{g_{a,b}^{\uparrow,1,A}(\Delta)
g_{b,a}^{\uparrow,2,A}(\Delta)}  \right\}
+(h_{ex} \rightarrow -h_{ex})
,
\end{equation}
where the overline is a disorder averaging, and where the
sites $a$ and $b$ belongs to the left and right interfaces
respectively (see
Fig.~\ref{fig:diag}a).
To obtain Eq.~(\ref{eq:IS-loc}), we start from
Eq.~(\ref{eq:sp-I-def}), use several times the
Dyson equation, and replace the Green's functions
of the isolated ferromagnets and superconductors
by Eqs.~(\ref{eq:g11-ferro}), (\ref{eq:g22-ferro})
and (\ref{eq:g-nonloc-S}). The integral over energy
is then calculated by contour integration. The poles
of the products of the anomalous Green's functions are
at $\omega=\Delta$, so that the ferromagnet
Green's functions are evaluated at $\omega=\Delta$
in Eq.~(\ref{eq:IS-loc}).

The spectral ``non local'' supercurrent involves
a diagram with propagation
in both ferromagnets (see Fig.~\ref{fig:diag}-(b)):
\begin{eqnarray}
\nonumber
I_S^{({\rm non loc})}(\omega)
&=&2\pi \frac{e}{h}\Delta |t_{a,\alpha}|^2
|t_{b,\beta}|^2  \sin{\varphi}
\mbox{Re} \left[
\sum_{a,b,a',b'}
\left\{ f_{\alpha,\beta}(\omega)
f_{\alpha',\beta'}(\omega)\right.\right.\\
&&\times \left. \left.
\left[ g_{a,b}^{\uparrow,1,A}(\omega)
g_{b',a'}^{\uparrow,2,A}(\omega) +
g_{a,b}^{\uparrow,2,A}(\omega) g_{b',a'}^{\uparrow,1,A}(\omega)
\right]\right\}  \right]\\
&+& (h_{ex} \rightarrow -h_{ex})
,\label{eq:IS-non-loc}
\end{eqnarray}
where the  sites $a$, $b$, $a'$ and $b'$
belong to different
interfaces (see Fig.~\ref{fig:diag}-(b)).
The variance of the non local supercurrent is obtained by integrating
the square of the spectral supercurrent given by Eq.~(\ref{eq:IS-non-loc})
over energy and averaging over disorder:
\begin{equation}
\overline{\left(I_S^{({\rm non loc})}\right)^2}
= \int d\omega 
\overline{\left(I_S^{({\rm non loc})}(\omega)\right)^2}
.
\end{equation}

\section{Evaluation of the average local supercurrent}
\label{sec:av-loc-super}

The
average over 
disorder of the product $\overline{g_{a,b}^{\uparrow,1,A}
g_{b,a}^{\uparrow,2,A}}$
in Eq.~(\ref{eq:IS-loc}), is evaluated
in the ladder approximation 
in the Appendix.
We find \cite{Melin05}
\begin{equation}
\overline{g_{a,b}^{1,1,A} g_{b,a}^{2,2,A}}
=- \frac{(\pi \rho_F)^2}{k_F^2 l_d d_{a,b}}
\exp{(i K_h d_{a,b})}
\exp{(-d_{a,b}/\xi_h)} 
,
\end{equation}
where
$d_{a,b}$ is the distance between the sites ``a'' and ``b''.
The decay length of the supercurrent of a single SFS junction
in the diffusive limit
is given by \cite{Melin05}
\begin{eqnarray}
\label{eq:xi-eff-1}
{1\over\xi_h} &=& \sqrt{\frac{3}{2 l_d}}
\sqrt{ \sqrt{\left(\frac{2}{l_\varphi^{({\rm ball})}}\right)^2
+(\Delta k)^2}+\frac{2}{l_\varphi^{({\rm ball})}}}
,
\end{eqnarray}
where the exchange field enters through $\Delta k = k_F^\uparrow
-k_F^\downarrow$, equal to the difference
between the spin-up and spin-down Fermi wave vectors,
with $\Delta k = 2 h / v_F$.
The wave vector of the supercurrent
oscillations\cite{Buzdin,Radovic,Heikkila,Golubov-short,osc,Aarts,Rya,Kontos,Guichard,Sellier} is given by
\begin{equation}
\label{eq:K-eff-1}
K_h = \sqrt{\frac{3}{2 l_d}}
\sqrt{ \sqrt{\left(\frac{2}{l_\varphi^{({\rm ball})}}\right)^2
+(\Delta k)^2}-\frac{2}{l_\varphi^{({\rm ball})}}}
,
\end{equation}
where $l_d$ is the elastic mean free path.
Due to the exchange field, the decay length $\xi_h$
given by Eq.~(\ref{eq:xi-eff-1}) is smaller
than the phase coherence length
\begin{equation}
l_\varphi=\sqrt{\frac{l_d l_\varphi^{({\rm ball})}}{3}}
.
\end{equation}
Eqs.~(\ref{eq:xi-eff-1}) and (\ref{eq:K-eff-1}) follow from the identity
\begin{equation}
\label{eq:xi-gene}
\frac{1}{\xi_h}+i K_h
=\sqrt{\frac{3}{l_d} \left(\frac{2}{l_\varphi^{({\rm ball})}}
+i \Delta k \right)}
,
\end{equation}
demonstrated in the Appendix.

After summing over all pairs of sites $(a,b)$ at the interface,
as detailed in the Appendix, we obtain
\begin{eqnarray}
\label{eq:local-super-av}
I_S^{({\rm loc})}&=&8\pi {e\over h} N_{\rm ch} \Delta |t_{a,\alpha}|^2
|t_{b,\beta}|^2 \frac{\pi^4 \rho_S^2 \rho_F^2}
{k_F^2 l_d a_0^2}
\frac{1}{\sqrt{(1/\xi_h)^2 +
K_h^2}}\\
&&
\exp{\left(-R/\xi_h\right)}
\cos{\left(K_h R-\theta\right)}
,
\nonumber
\end{eqnarray}
with $\tan{\theta}=K_h/\xi_h$.
The local supercurrent is then
proportional to the number of channels
$N_{\rm ch}$ at a single interface, $N_{\rm ch}$ being itself proportional
to the area of a single junction.

\section{Statistical fluctuations of the non local supercurrent}
\label{sec:fluc-super}
The average of the non local supercurrent vanishes because of disorder
averaging in the diffusive system\cite{Melin03}, or because of
averaging over the Fermi oscillations in the ballistic
system. 
The disorder average of the square of the 
non local supercurrent 
$\overline{(I_S^{({\rm non loc})})^2}$
involves the diagrams on Fig.~\ref{fig:diag}-(c), the
average over disorder of which does not decay exponentially over the
elastic mean free path $l_d$.

The non local supercurrent (\ref{eq:IS-non-loc}) can be recast in
the form
\begin{equation}
\label{eq:I-S-non-loc-X}
I_S^{({\rm non loc})}= {\cal A} \sum_{a,b,a',b'} \sum_{\sigma,\tau}
\left( X_{a,b,a',b'}^{(\sigma,\tau),A} + 
X_{a,b,a',b'}^{(\sigma,\tau),R} \right)
,
\end{equation}
where ${\cal A}=\pi (e/h) \Delta |t_{a,\alpha}|^2
|t_{b,\beta}|^2$, and
$X_{a,b,a',b'}^{(\sigma,\tau)}=f_{\alpha,\beta}
f_{\alpha',\beta'}
g_{a,b}^{(\sigma,\tau)}
g_{b',a'}^{(\sigma,\overline{\tau})}$,
and where $\sigma=\uparrow,\downarrow$
is the projection of the spin along the $z$ axis, and
$\tau=1,2$ is the Nambu index. 
The notation $\overline{\tau}$ in
the definition of $X_{a,b,a',b'}^{(\sigma,\tau)}$ 
corresponds to
$\overline{\tau}=1$ if $\tau=2$, and $\overline{\tau}=2$ if
$\tau=1$. We deduce from Eq.~(\ref{eq:I-S-non-loc-X})
\begin{equation}
\label{eq:XX}
\overline{ \left( I_S^{({\rm non loc})} \right)^2} =
2 {\cal A}^2 \sum_{a,b,a',b'} \sum_{\sigma,\tau} \sum_{\sigma',\tau'}
\mbox{Re} \left[ 
\overline{ X_{a,b,a',b'}^{(\sigma,\tau),A} X_{a,b,a',b'}^{(\sigma',\tau'),A}}
+ 
\overline{ X_{a,b,a',b'}^{(\sigma,\tau),A} X_{a,b,a',b'}^{(\sigma',\tau'),R}}
\right]
,
\end{equation}
corresponding to the diagrams on Fig.~\ref{fig:diag}-(c).
The disorder averages in each electrode are then carried out,
following the Appendix.
Factoring out the propagators in the superconductor\cite{note1},
and carrying out the summation over the conduction channels, leads to
\begin{eqnarray}
\nonumber
\overline{ \left(I_S^{\rm non loc}\right)^2}
&=&4 \pi^2 (e/h)^2 \Delta^2 |t_{a,\alpha}|^4
|t_{b,\beta}|^4 N_{\rm ch}^2
\overline{f_{\alpha,\beta}^2}\; \overline{f_{\alpha',\beta'}^2}
\left(\frac{(\pi \rho_S)^2 l_\varphi }
{(k_F a_0)^2 l_d}\right)^2\\
&\times&
\left\{  \exp{\left(-\frac{2 R}{l_\varphi}\right)} +
\frac{1}{\sqrt{1+(l_\varphi (k^\uparrow-k^\downarrow))^2}}
\exp{\left(-\frac{2 R}{\xi_h}\right)}
\right\} \sin^2{\varphi}
\label{eq:moment2-nonloc}
,
\end{eqnarray}
where we discarded the terms decaying exponentially over the Fermi
wave-length. 
The product $\overline{f_{\alpha,\beta}^2}
\overline{(f'_{\alpha,\beta})^2}$ is proportional to
$1/D^2$, with $D^2$ of order $N_{\rm ch} a_0^2 (D/r)^2$, so that
$\overline{ \left(I_S^{\rm non loc}\right)^2}$ scales like
$N_{\rm ch}$ (see Fig.~\ref{fig:schema-coupe} for the
notations $D$ and $r$).
The variance of the supercurrent given by
Eq.~(\ref{eq:moment2-nonloc}) involves a ``long range'' contribution
decaying over $l_\varphi$, and a short range contribution
decaying over $\xi_h$. The former propagates over a much
larger distance than the latter. Both contributions are
identical for normal metals, but the long range contribution 
dominates for ferromagnets with $\xi_h \alt R \alt l_\varphi$.

\section{Conclusions}
\label{sec:conclu}

To conclude, we have investigated the possibility of coupling coherently
two superconductors by two spatially separated ferromagnets. The
statistical fluctuations of the
Josephson current are proportional to the square root of the surface of the
junctions. The fluctuating part of the Josephson current is ``long range''
in the sense that it does not decay over the exchange length
$\xi_h$, but decays over the phase coherence length $l_\varphi$.
We predict a Josephson current mediated by fluctuations if the
length of the ferromagnets is larger than $\xi_h$ and smaller than
$l_\varphi$.
The supercurrent is expected to fluctuate as a function
of the relative spin orientation of the ferromagnets.
This effect
can be used as a test of the phase coherence of crossed Andreev reflection,
without the competition between the crossed Andreev reflection and
elastic cotunneling channels\cite{Falci01,Melin04} since the Josephson
effect probes solely the anomalous propagator in the superconductor.

Another proposal has been made recently to probe
a long range Josephson effect in ferromagnets with non collinear
magnetizations in a single SFS junction,
generating triplet correlations\cite{Bergeret,Kadi}
that can also propagate up to $l_\varphi$. This effect is not
equivalent to the one considered here since it involves
propagation in a single electrode. The fluctuations of the
Josephson current discussed here are also not equivalent to
universal conductance fluctuations\cite{Lee} since the root mean square 
of the supercurrent distribution is proportional to the
square root of the junction area.

\section*{Acknowledgments}
The author acknowledges fruitful discussions with
D. Feinberg, and thanks H. Courtois for
useful comments on the manuscript. 

\appendix

\section{Details of disorder averaging}
\label{app:des}

\subsection{Elastic scattering time}
The Green's function $\hat{G}^{(0)}({\bf k},\omega)$
of a disordered isolated ferromagnet, diagonal in the
Nambu representation, is given by
$\hat{G}^{(0)}({\bf k},\omega)=\hat{g}({\bf k},\omega)
\left[ \hat{I} + \hat{\Sigma}_v \hat{G}^{(0)}({\bf k},\omega)
\right]$,
where $\hat{g}({\bf k},\omega)$ is the Green's function
of a ballistic isolated ferromagnet, and
\begin{equation}
\label{eq:Sigma-v}
\hat{\Sigma}_v=n v^2 \int \frac{d {\bf k'}}{(2\pi)^3}
\hat{g}({\bf k}',\omega)
,
\end{equation}
is the disorder self-energy, where $n$ is the concentration of
impurities. Evaluating the integral in Eq.~(\ref{eq:Sigma-v})
by contour integration leads to
\begin{eqnarray}
\hat{G}^{(0),\uparrow,1,A}({\bf k},\omega)&=&
\frac{1}{\omega-h-\xi_k - i /\tau_{1,1}}\\
\hat{G}^{(0),\uparrow,2,A}({\bf k},\omega)&=&
\frac{1}{\omega-h+\xi_k - i /\tau_{2,2}}
,
\end{eqnarray}
with
\begin{eqnarray}
\tau_{1,1}&=&\frac{\tau}{1+\omega/2\epsilon_F-h/2\epsilon_F}\\
\tau_{2,2}&=&\frac{\tau}{1-\omega/2\epsilon_F+h/2\epsilon_F}
,
\end{eqnarray}
with $\tau=4\pi \epsilon_F/(k_F^3 v^2)$ the elastic scattering
time with $h=\omega=0$.

\subsection{Disorder averaging of the supercurrent}
The supercurrent involves the disorder average $\overline{
g_{a,b}^{\uparrow,1,A}(\omega) g_{b,a}^{\uparrow,2,A}(\omega)}$,
evaluated in Fourier space in the ladder approximation \cite{Smith}.
Using contour integration,
we find 
\begin{equation}
\int \frac{d {\bf k}}{(2\pi)^3}
\hat{G}^{\uparrow,1,A}({\bf k},\omega)
\hat{G}^{\uparrow,2,A}({\bf k}+{\bf q},\omega)
\simeq -1+i\tau (\omega-h+i v_F/l_\varphi^{(\rm ball)}) 
+ \frac{q^2 \tau^2 \epsilon_F^2}
{3 k_F^2}
.
\end{equation}
After the summing the ladder diagrams, the poles are found
at wave-vector
\begin{equation}
\label{eq:q}
q^{({\rm diff})}=\sqrt{\frac{6}{l_d} \left(\frac{1}{l_\varphi^{({\rm ball})}}
-\frac{i h}{v_F}+\frac{i}{\xi^{({\rm ball})}}\right)}
,
\end{equation}
where we replaced $\omega$ by $\Delta$, as obtained in the
energy integration of the local supercurrent [see Eq.~(\ref{eq:IS-loc})].
Eq.~(\ref{eq:q})
leads directly to Eq.~(\ref{eq:xi-gene}) if one assumes
the ballistic
superconducting coherence length $\xi^{({\rm ball})}$
of the order of $1 \, \mu$m
to be large compared to the ballistic exchange length $v_F/h$
and to the ferromagnet ballistic phase coherence length
$l_\varphi^{({\rm ball})}$.
The geometrical prefactor
$(\pi \rho_F)^2/k_F^2 l_e d_{a,b}$
is then obtained by evaluating the
residue in the
integral over ${\bf q}$.

\subsection{Summation over the conduction channels}
\label{app:sum}
The supercurrent given by Eq.~(\ref{eq:IS-loc}) involves a
summation over the conduction channels. This summation is
evaluated through
\begin{eqnarray}
\nonumber
\sum_{a,b} \frac{a_0}{d_{a,b}}
\exp{\left[-\left(\frac{1}{\xi}-iK_h\right)d_{a,b}\right]}
&\simeq& N_{\rm ch} \int
\frac{2\pi y dy}{a_0 \sqrt{R^2+y^2}}
\exp{\left[-\left(\frac{1}{\xi}-iK_h\right)
\sqrt{R^2+y^2}\right]}\\
&=&N_{\rm ch} \frac{\xi/a_0}{1-i K_h \xi}
\exp{\left[-\left(\frac{1}{\xi}-iK_h\right)R\right]}
,
\end{eqnarray}
where $d_{a,b}=\sqrt{R^2+y^2}$ is the distance between the
sites $a$ and $b$.

\subsection{Disorder averaging from real space Green's functions}

The product of the ballistic Green's functions
$g_{a,b}^{\uparrow,1,A}(\omega) g_{b,a}^{\uparrow,2,A}(\omega)$
is proportional to
$\exp{[-q^{({\rm ball})} d_{a,b})]}$, with
$q^{({\rm ball})}= i\Delta k +2 i\omega / v_F+
2/l_\varphi^{({\rm ball})}$.
The wave-vector
$q^{({\rm diff})}$
given by Eq.~(\ref{eq:q}) is related to
$q^{({\rm ball})}$ according to
\begin{equation}
\label{eq:q-b-d}
q^{({\rm diff})} = \sqrt{\frac{3 q^{({\rm ball})}}{l_d}}
.
\end{equation}


\begin{thebibliography}{99}
\bibitem{note1} This asumption is stricktly speaking
  valid only for $D \gg r$ (see 
  Fig.~\ref{fig:schema-coupe}). However, Eq.~(\ref{eq:moment2-nonloc})
  is still valid for $D\sim r$, up to a dimensionless
  geometrical prefactor.
\end{thebibliography}
\end{document}